\documentclass[sigconf]{acmart}
\usepackage[nolist]{acronym}
\usepackage{hyperref}
\usepackage{tabularx}
\usepackage{xcolor}
\usepackage{cleveref}
\usepackage[most]{tcolorbox}
\usetikzlibrary{calc,shadows.blur}
\tcbuselibrary{skins}

\newcommand{\fix}[1]{\textcolor{black}{ #1}}

\newtcolorbox{resp}[1][]{%
	enhanced jigsaw,
	colback=gray!20!white,%
	colframe=gray!80!black,
	size=small,
	boxrule=0pt,
	arc=0pt,
	outer arc=0pt,
	frame hidden,
	drop shadow=black!50!white
}

\newtcolorbox{resp_float}[1][]{%
	enhanced jigsaw,
	colback=gray!20!white,%
	colframe=gray!80!black,
	size=small,
	boxrule=0pt,
	arc=0pt,
	outer arc=0pt,
	frame hidden,
	drop shadow=black!50!white,
	float
}

\AtBeginDocument{%
\providecommand\BibTeX{{%
\normalfont B\kern-0.5em{\scshape i\kern-0.25em b}\kern-0.8em\TeX}}}

\copyrightyear{2020} 
\acmYear{2020} 
\setcopyright{acmlicensed}\acmConference[AutomotiveUI '20]{12th International Conference on Automotive User Interfaces and Interactive Vehicular Applications}{September 21--22, 2020}{Virtual Event, DC, USA}
\acmBooktitle{12th International Conference on Automotive User Interfaces and Interactive Vehicular Applications (AutomotiveUI '20), September 21--22, 2020, Virtual Event, DC, USA}
\acmPrice{15.00}
\acmDOI{10.1145/3409120.3410638}
\acmISBN{978-1-4503-8065-2/20/09}

\begin{document}

\begin{acronym}[]
	\acro{ECU}{Electronic Control Unit}
	\acro{GDPR}{General Data Protection Regulation}
	\acro{HCI}{Human-Computer Interaction}
	\acro{HMI}{Human-Machine Interaction}
	\acro{IVIS}{In-Vehicle Information System}
	\acro{KPI}{Key Performance Indicator}
	\acro{OEM}{Original Equipment Manufacturer}
	\acro{RH}{Research Hub}
	\acro{UCD}{User-centered Design}
	\acro{UX}{User Experience}
\end{acronym}

%
\title[Field User Interaction Data in the Automotive UX Development]{The Role and Potentials of Field User Interaction Data in the Automotive UX Development Lifecycle: An Industry Perspective}

%
\author{Patrick Ebel}
\email{patrick.ebel@tu-berlin.de}
\orcid{0000-0002-4437-2821}
\affiliation{%
	\institution{Technische Universit{\"a}t Berlin}
	\city{Berlin}
	\state{Germany}
}
\author{Florian Brokhausen}
\email{florian.brokhausen@tu-berlin.de}
\orcid{0000-0002-8003-4344}
\affiliation{%
	\institution{Technische Universit{\"a}t Berlin}
	\city{Berlin}
	\state{Germany}
}
\author{Andreas Vogelsang}
\email{andreas.vogelsang@tu-berlin.de}
\orcid{0000-0003-1041-0815}
\affiliation{%
	\institution{Technische Universit{\"a}t Berlin}
	\city{Berlin}
	\state{Germany}
}

%

\begin{abstract}
We are interested in the role of field user interaction data in the development of \acp{IVIS}, the potentials practitioners see in analyzing this data, the concerns they share, and how this compares to companies with digital products.
We conducted interviews with 14 UX professionals, 8 from automotive and 6 from digital companies, and analyzed the results by emergent thematic coding. Our key findings indicate that implicit feedback through field user interaction data is currently not evident in the automotive UX development process. Most decisions regarding the design of \acp{IVIS} are made based on personal preferences and the intuitions of stakeholders. However, the interviewees also indicated that user interaction data has the potential to lower the influence of guesswork and assumptions in the UX design process and can help to make the UX development lifecycle more evidence-based and user-centered.
\end{abstract}

%
%

%
%

%
\maketitle

\section{Introduction}
The influence of software-based systems on the in-car \ac{UX} has changed from purely operating the car, through adding simple infotainment devices to the highly complex systems we experience today~\cite{Harvey.2016}. Modern cars are equipped with various \acfp{IVIS} that offer a large variety of features, interaction possibilities, and can be controlled via multiple interfaces~\cite{Lamm.2019}. The expectations towards those systems are not only driven by the customer's experiences with other car manufacturers but also by modern smartphones, websites, and other digital products. \citeauthor{Baker.2016}~\cite{Baker.2016} report that customers are willing to pay up to $15\,\%$ of a car's list price, or as much as \$10,000 for connected car and infotainment technology. This paradigm shift leads to a growing and diverse competition in the automotive domain and unlocks potential for advanced and intelligent in-vehicle features~\cite{Pfleging.2012, Schroeter.2012, Ebel.2020}. As a consequence, the task of designing user interfaces that meet customer demands has become more challenging and represents a crucial part of automotive development~\cite{Harvey.2016}.

\ac{UCD} requires experienced designers and exhaustive user studies and, therefore, is a costly, yet critical, task. This applies in particular to the design and evaluation of \acp{IVIS}. Compared to many digital products, the complex automotive architecture, the critical requirements regarding functional safety, and the traditional structures in the automotive industry are challenging~\cite{Broy.2006, Alvarez.2017}.
Additionally, the driver's experience does not only depend on the system itself but also on the current driving and traffic situation~\cite{Schneegass.2013,Ahmad.2015}. While driving a car, the interaction with the \ac{IVIS} is only a secondary task, which makes the interactions highly context-sensitive \cite{Harvey.2016}. Therefore, it is necessary to create a realistic simulation of the driving situation when evaluating \acp{IVIS}. Simulating the driving situation provides value to the designers but requires high investments. While more cost-effective alternatives~\cite{Gerber.2019} are developed, they are still in an early research stage. The analysis of field user interaction data can reduce the effort for extensive user studies and expensive simulations by gaining insights about user behavior through data analysis~\cite{Orlovska.2018}.

Despite the generally growing awareness of the potentials of big data analysis, there is a lack of research on how data-driven approaches may support the automotive UX development process. In this paper, we present results from a qualitative study on the current role of field user interaction data in the automotive industry and highlight the differences to digital products. 
We conducted semi-structured interviews with 14 \ac{UX} professionals, where 8 are currently employed in the automotive industry and 6 in other industries. In the interviews, we addressed the current state-of-the-art, the challenges, and the potentials of field user interaction data in the respective \ac{UX} development lifecycles. 
Our key findings indicate that implicit feedback through field user interaction data is currently not evident in the automotive UX development process.
However, practitioners confirm its potential to make the UX development lifecycle more evidence-based and user-centered. Additionally, participants are concerned about insurmountable organizational, legal, or technical restrictions that prevent data collection. Participants from digital companies share most of the needs and potentials with the automotive participants but are generally concerned with more advanced issues like data interpretability and triangulation between qualitative and quantitative feedback.

\section{Background}
In this section, we discuss the concept of \ac{UX}, its different perception in industry and academia, and its special role in the automotive industry. We present a definition of field user interaction data and give an overview of related studies.
	
\subsection{UX and its Role in the Automotive Industry}

A good \ac{UX} is the main goal of most product development processes. However, \ac{UX} is perceived differently in academia and industry. Academics commonly agree that UX is a holistic and subjective concept~\cite{Roto.2009}, which goes beyond solving usability problems or creating a pleasant UI. Usability aspects contribute to the overall UX but do not suffice as stand-alone UX measures since they do not yield insights on how the interaction was perceived by the user~\cite{Roto.2011}. The perceived UX is mainly influenced by the user's internal state, the characteristics of the system at use, and the context in which the interaction occurs~\cite{Hassenzahl.2006}.

This holistic concept of UX leads to challenges in practice \cite{Gray.2014, Vaananen.2008} with some of them being particularly relevant in automotive UX development. On the one hand, UX is strongly affected by the context in which the interaction occurs \cite{Harvey.2011, Orlovska.2020}. The interaction with an \ac{IVIS} is therefore dependent on environmental conditions, such as the driving scenario, the dual-task environment, and the frequency of use \cite{Harvey.2010}. Those aspects must be addressed by the designers, increasing the complexity of the design task \cite{Fastrez.2008} \fix{by the need for a context-dependent presentation of information \cite{Loecken.2017}}. On the other hand, practitioners and organizations are more interested in the long-term experience rather than in temporary emotions~\cite{Law.2009}. Regarding the automotive industry with its long product lifecycles~\cite{Broy.2006}, its different touch-points, \fix{and diverse and global user base \cite{Heimgaertner.2017}} the question of how to capture an `overall' UX score~\cite{Law.2009} gets even more interesting. Additionally, \citeauthor{Frison.2019b} emphasize the importance of UX, as it influences trust in the vehicle~\cite{Frison.2019b} and gets even more important with the transition toward automated driving~\cite{Frison.2019}.

\subsection{Field User Interaction Data}

In line with the definition of \citeauthor{Harvey.2011} \cite{Harvey.2011}, we define user interaction data as every kind of interaction between the user and the \ac{IVIS}. User interactions are neither bound to a specific type of interface nor a specific type of interaction and therefore vary in their granularity. User Interaction data can be generated both in the field and in the lab. In the following, we will mainly focus on user interaction data generated during field usage. \fix{We define this as \emph{field user interaction data}, which is continuously and implicitly recorded automatically in all production line vehicles. }

\subsection{Related Work} \label{subsec:rel_work}
The approach of using field user interaction data to evaluate and enhance a product's UX is already well researched and widely established in digital domains like web and app development. In 2006, \citeauthor{Atterer.2006}~\cite{Atterer.2006} presented an approach for detailed user tracking on web pages that benefits usability evaluation by incorporating implicit user interactions. While \citeauthor{Atterer.2006} track multiple different user actions, \citeauthor{Navalpakkam.2012}~\cite{Navalpakkam.2012} propose an approach that predicts the overall experience of web page users by mouse tracking alone. \citeauthor{Nebeling.2013}~\cite{Nebeling.2013} developed a framework that combines automated usability testing with crowdsourcing. They argue that the benefits of large-scale online user testing outweigh the trade-offs compared to lab studies. Compared to digital domains, the research field on the usage of implicit interaction data to evaluate \acp{IVIS} is not yet widely explored.

This aligns with the findings of \citeauthor{Lamm.2019}~\cite{Lamm.2019}, who found that model-based approaches and automated evaluations do not play an important role in automotive UX development. However, they do not reflect on the views of practitioners and what might prevent them from applying these approaches.

A method to measure UX in an automotive context based on the fulfillment of psychological needs is proposed by \citeauthor{Korber.2013b}~\cite{Korber.2013b, Korber.2013}. They argue that the presented questionnaire is not only able to measure the UX and compare products quantitatively but also to predict possible experiences in early prototype stages. 

Another approach is presented by \citeauthor{Harvey.2011}~\cite{Harvey.2011}. They propose a framework based on thirteen methods that measure the objective and subjective levels of performance and workload of users interacting with \ac{IVIS}. \fix{Similarly, \citeauthor{Alvarez.2017}~\cite{Alvarez.2017} present a set of tools and methodologies that aim to benefit the creation of a holistic automotive design space. The work of \citeauthor{Riener.2017}~\cite{Riener.2017} gives an extensive overview on how the driver-interaction loop can be modeled.} Although some of the presented methods are based on user interaction data, the use of data, implicitly \fix{, automatically, and continuously} collected from \fix{the whole user base}, is not considered.

This is investigated by \citeauthor{Orlovska.2018}~\cite{Orlovska.2018}, showing how field user interaction data can support the evaluation of \acp{IVIS}. They argue that field data-driven approaches enable more accurate feedback and enable testing of the whole user base. They also found that software platforms of \acp{OEM} are not designed for user behavior logging and that methods need to be developed that enable compliance with data protection regulations. Additionally, they present a case study~\cite{Orlovska.2018b} on how field user interaction data can benefit the overall usability assessment. \fix{However, this study exclusively deals with a specific system of a single manufacturer. Therefore, the results are hard to generalize.}

\section{Study Design}

Despite the claimed potentials of using field user interaction data to improve the UX development lifecycle and its success in other fields, there are indications that these potentials are not (yet) leveraged in the development of automotive \acp{IVIS}. We are interested in \emph{why} this is the case.
We want to answer the following research questions:
\begin{itemize}
	\setlength{\itemindent}{+.0in}
	\item[\textbf{RQ1:}] What is the current role of field user interaction data in the automotive UX development lifecycle?
	\item[\textbf{RQ2:}] What are the needs, challenges, and concerns in the context of data-driven UX Development?
	\item[\textbf{RQ3:}] How can the automotive UX development lifecycle benefit from field-data-driven approaches?
	\item[\textbf{RQ4:}] What is specific to the automotive UX development lifecycle and what can be generalized from digital companies?
\end{itemize}

\begin{table*}
	\caption{Participants}
	\centering
	\label{tab:participant}
	\small
	\begin{tabular}{@{}ccccccc@{}}
		\toprule
		\# & Age &  Education & Job Title & Industry &	\# Employees & Professional Experience\\
		\midrule
		P1 	& 20-29 	& Master 		& User Researcher & Automotive (RH)	&  201-500		& 1-4 years\\
		P2 	& 40-49 	& PhD & Technical Specialist & Automotive (OEM)	& 10,001-100,000 	& 10-19 years \\
		P3 	& 20-29 	& Bachelor 		& UX/UI Designer  & Automotive (RH)	& 51-201			& 1-4 years \\
		P4 	& 30-39 	& Master 		& UX/UI Designer & Automotive (RH)		&  201-500		& 1-4 years  \\
		P5 	& 30-39 	& Bachelor 		& UX/UI Designer  & Automotive (RH)	& 201-500			& 10-19 years \\
		P6 	& 30-39 	& Master 		& UX Marketing Specialist  & Automotive (OEM)	& 10,001-100,000		& 1-4 years \\
		P7 	& 40-49 	& Bachelor 		& Interaction Designer  & Automotive (RH) 	& 501-1,000			& 10-19 years\\
		P8 & 40-49 		& Diploma 		& Project Manager UX  & Automotive (OEM)		&  10,001-100,000			& 1-4 years \\
		P9 	& 20-29 	& Master 		& Interaction Designer 	 & Internet of Things & 501-1,000			& 5-9 years \\
		P10 & 30-39  	& PhD & UX Manager 			 & E-Commerce & 10,001-100,000 	& 5-9 years \\
		P11 & 30-39 	& PhD & Ergonomist 			 & Telecommunications  & 100,000+ 			& 1-4 years \\
		P12 & 40-49 	& Diploma 		& Head of UX 		& IT Service Provider	& 1,001-10,000 		& 10-19 years \\
		P13 & 40-49 	& Bachelor 		& UX/UI Designer  & Apps and Software		&  51-200			& 1-4 years \\
		P14 & 30-39 	& Master 		& Design Manager  & Digital Music Service	&  1001-10,000		& 10-19 years \\	
		\bottomrule
	\end{tabular}
\end{table*}

\subsection{Research Method and Interview Design}

To answer the research questions we followed a qualitative approach and conducted semi-structured interviews. Before conducting the interviews, we asked the participants to answer a questionnaire regarding their demographics, background, and experience. Although we prepared a list of questions\footnote{Interview Guideline: \url{https://figshare.com/s/d9af6f2fa18f59e7e7eb}}, we varied the order of questions to unfold the interview conversationally. This exploratory approach allows open-ended questions and engages the participants to independently address the objectives they consider important. The interview itself was divided into three parts addressing the three usability engineering lifecycle phases introduced by \citeauthor{Nielsen.1992}~\cite{Nielsen.1992}: pre-design, design, and post-design. Regarding each phase, we asked the participants about the methods they currently apply, the challenges they face, and the potentials they see in data-driven approaches. Each interview lasted approximately one hour \fix{and was conducted by the first two authors with always one interviewee present. Of the 14 interviews, 5 were carried out in person, one via video call and 8 via phone.}

\subsection{Study Subjects}

In total, we interviewed 14 \ac{UX} professionals from 11 different companies, 8 working in the automotive industry, and 6 working for digital companies. We define a digital company as a company whose main product is a digital product or which has a digital product in its core business.
\fix{The domains of these digital companies range from digital music services through e-commerce to telecommunications. However, we carefully selected candidates that are solely responsible for a digital product within their company. To get this broad range} of perspectives inside each of the groups we applied purposive sampling~\cite{Etikan.2016}. We approached companies of different sizes and domains and selected candidates of various backgrounds. The interviews were conducted \fix{between October 2019 and March 2020.} Since all participants are kept anonymous, they are referred to with IDs P1-P14. All participants are currently employed in industry, with only 3 never having worked in a research context. \Cref{tab:participant} shows an overview of the demographics of the participants. In the automotive industry, it is very common that \acp{OEM} have multiple smaller research facilities or \acp{RH}, where specialists work on a specific topic, decoupled from the main company. The participants did not receive any compensation.

\subsection{Data Analysis}
The first author transcribed and anonymized the audio recordings of the interviews. Afterward, the first and second authors applied a mixture of a priori and emergent coding~\cite{Saldana.2015} in a collaborative manner using ATLAS.ti\footnote{\url{https://atlasti.com/}}.

For initial coding, both authors agreed on a set of codes based on the research questions. However, the authors were free to introduce new codes whenever they considered it to be necessary. For the coding, no special restrictions applied and each interview transcript was coded independently by the first two authors. To ensure the reliability of coding, the inter-coder agreement, according to \citeauthor{Krippendorff.2018}~\cite{Krippendorff.2018}, was calculated before the results of each interview were discussed and merged. The inter-coder agreement over all interviews is $\alpha=0.822\,(\sigma=0.119)$ representing a satisfactory result \cite{Krippendorff.2018}.
Newly introduced codes were reviewed by both authors and after mutual agreement, were added to the set of codes. This procedure was repeated for each interview and already coded transcripts were updated collectively by the authors. The changes introduced to the set of codes decreased after 6 interviews and no new codes emerged after 11 interviews. Therefore, we conclude further interviews will provide only a few (if any) new insights and we reached a point of \emph{theoretical saturation}~\cite{Guest.2006}.

The quotes from non-English speaking interviewees were translated into English and edited for readability. Colloquial expressions were not changed to reflect the informal setting of the interview.

\begin{figure*}
	\centering
	\includegraphics[width=0.9\linewidth]{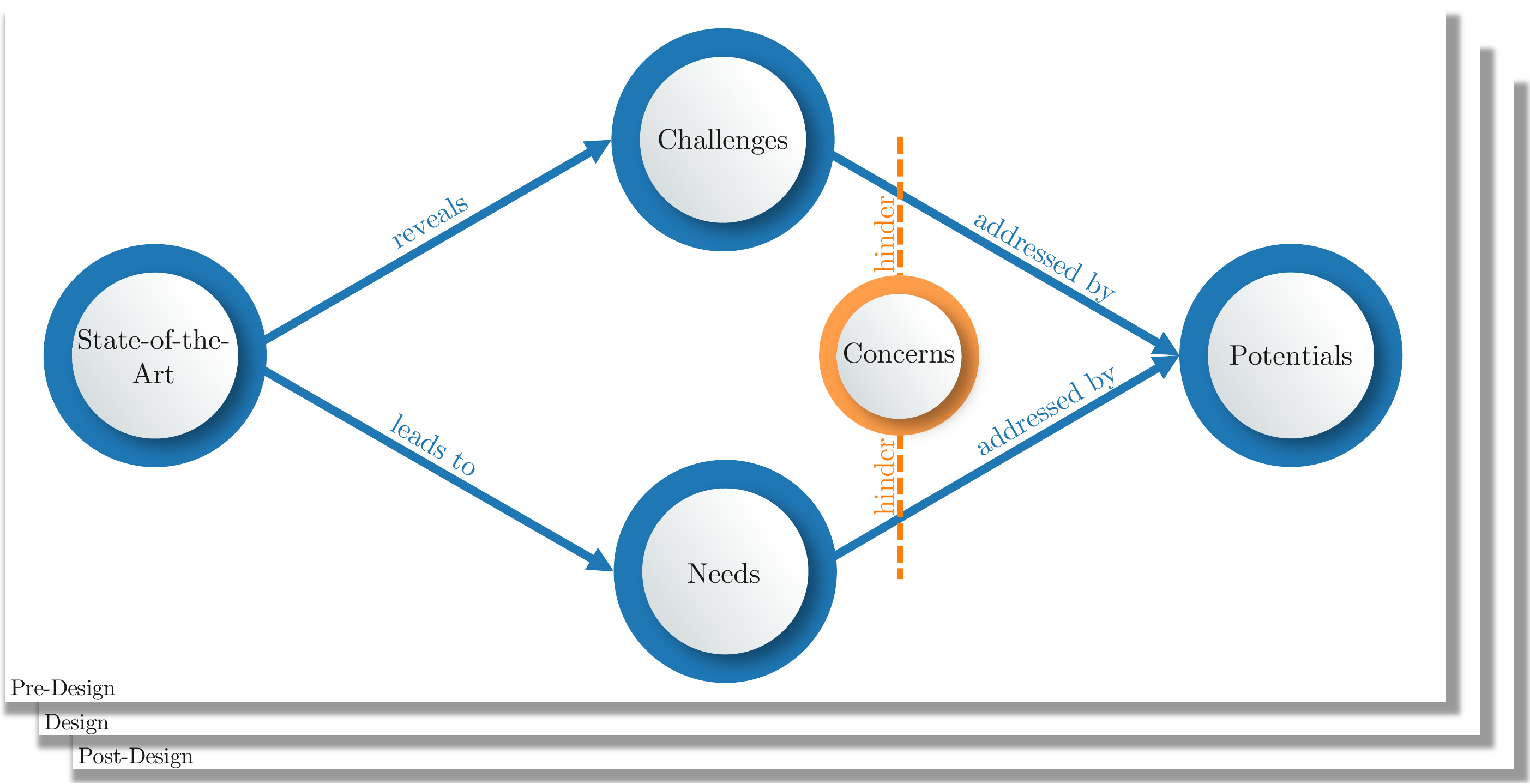}
	\caption{Thematic Coding Model}
	\label{fig:Model}
\end{figure*}

\subsection{Threats to Validity}
The five threats to validity in qualitative research identified by \citeauthor{Maxwell.2012}~\cite{Maxwell.2012} also apply to our study design. These threats describe the flaws that can occur while obtaining and interpreting the study observations. Further, the collected data might be manipulated to fit a specific theory, may it be deliberately or accidentally. \citeauthor{Maxwell.2012}~\cite{Maxwell.2012} argues that the researchers must preclude those threats by developing a study design that provides evidence that no ``alternative hypotheses'' can be made~\cite{Lewis.2009}.

\textit{Descriptive validity} refers to the threat of incomplete and inaccurate recordings. To preclude this threat all interviews were recorded and transcribed. The transcripts are annotated with timestamps such that the original conversation can be traced back during analysis. 

The threat of \textit{Interpretation validity} addresses the challenge to capture the observations as intended by the participants. To avoid this threat, we used open-ended and non-directional questions. Additionally, all interviews were independently coded by two authors, and potentially ambiguous statements were discussed to identify the interpretation intended by the participant. 

\textit{Theory validity} and \textit{researcher bias} refer to the threat that the researchers force the data to fit a certain theory they want to support or that they possess a deliberate bias regarding the participants or a certain outcome. Mitigating this threat is the fact that the study is constructed to be very exploratory, having the intention to reflect the current state-of-the-art in the industry and identify potentials. Additionally, we lowered the researcher bias by applying the introduced coding and reviewing concepts. 

\textit{Reactivity} describes the threat that the presence of the interviewers may influence the interviewees. This threat can hardly be mitigated but still, the authors payed attention to not influence the interviewees when conducting the interviews.

\fix{An additional threat is posed by the selection of the interviewees. We only interviewed employees of automotive \acp{OEM}, which might introduce some bias by excluding suppliers. The \ac{OEM} research hubs usually act as company-internal suppliers, being solely responsible for whole systems within the car, which might add some similar perspectives.}

\section{Results}

We structured the identified codes into categories and illustrate their relations in \Cref{fig:Model}.

\begin{table}[h]
	\centering
	\renewcommand{\arraystretch}{1.3}
	\begin{tabularx}{\linewidth}{lX} 
		\textbf{State-of-the-Art:}	& Statements of phenomena in current practice, which reveals a challenge or leads to a need\\
		\textbf{Challenges:}	& Statements of problems that arise from current practices\\
		\textbf{Needs:}	& Statements of demands towards improving the UX development lifecycle\\
		\textbf{Concerns:}	& Statements of doubts that a challenge can be overcome or a need can be fulfilled \\ 
		\textbf{Potentials:}	& Statements of areas where data-driven approaches may address a challenge or fulfill a need\\
	\end{tabularx}
\end{table}

The model shows that the reported \textbf{State-of-the-Art} reveals \textbf{Challenges} and leads to \textbf{Needs} of practitioners. Some of these challenges and needs can be addressed by analyzing field user interaction data (\textbf{Potentials}). These potentials are expressed explicitly and implicitly by the participants. \textbf{Concerns} were mentioned as hindering factors. The model applies to the pre-design, design, and post-design phase.

\begin{figure*}
	\centering
	\includegraphics[width=\linewidth]{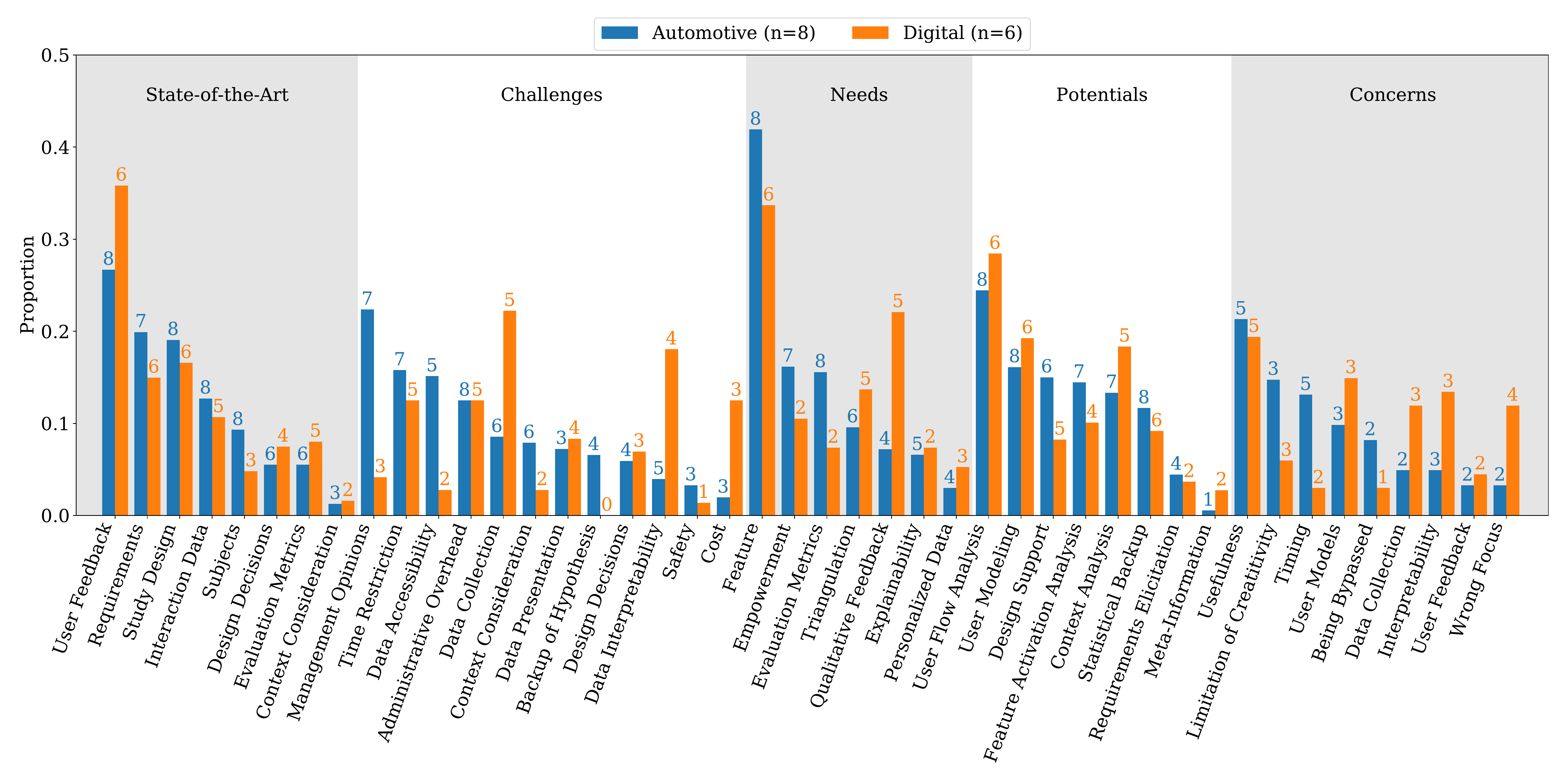}
	\caption{Code Distribution Graph}
	\label{fig:BarPlot}
\end{figure*}

\Cref{fig:BarPlot} shows the distribution of codes within each category as bars, and the number of interviews the code occurred in as numbers on top. On average we introduced 80 $(\sigma=26)$ codes per interview.

\subsection{State-of-the-Art}
In the pre-design phase, the most frequent code in the state-of-the-art category is \textbf{Requirements}. Six of the eight interviewees from the automotive domain agree that the requirement and feature elicitation is not user-oriented.
P1 states: \textit{``[\dots] at this point [the pre-design phase] we have no clue if the customer [user] is interested in this feature or not''}.
Only interviewee P2 from the automotive domain confirmed that they already use some form of field user interaction data in the pre-design stage by aggregating data from company cars in all markets in real-time.
This data is then used, for example, to derive statements about the usage frequency of certain features to \textit{``prioritize what [the company] actually should spend money on and what [\dots] the most important features based on usage [are]''} (P2).

Another quarter of citations in the pre-design phase are tagged as \textbf{User Feedback}, being the most mentioned state-of-the-art code over all phases (see \Cref{fig:BarPlot}). All participants from the automotive domain but P5 mention they receive some form of user feedback in the pre-design phase. This feedback, however, is usually in the form of general market research and not really focused on the users' explicit needs or behavior. In contrast, P2 describes their rather elaborate process of analyzing user needs: \textit{``We do that [long-term ethnographic research] by observing, interviewing, participating with people in their life, being in their homes, trying to figure out what life people are living, what are their pleasure points and pain points''}.
In contrast, within digital companies, the elicitation of features and requirements seems to be more focused on the user. All digital domain participants report that the ideas in early development phases are created together with representative customers, are based on insights drawn from field user interaction data, or both.

In the design phase, two-thirds of state-of-the-art citations are coded with \textbf{Design Decisions}, \textbf{Study Design}, or \textbf{User Feedback}.
Regarding the automotive state-of-the-art, 6 out of 8 practitioners confirm that they evaluate their designs mainly in-house in an informal, qualitative way with coworkers and other UX experts.
In contrast, P2 and P6 from the automotive domain confirm that they recruit external people on a regular basis for early feedback on their designs and ideas.
Interviewee P7 describes the current state-of-the-art: \textit{``Testing within [the company] is sort of ok if you just need to do something quickly, but if we want to verify things, it's better to get people that are not familiar with what we do''}.
While all participants implement clickable prototypes for their products, these are only evaluated qualitatively. The automotive participants agree that their current process of gathering qualitative feedback on prototypes is quite advanced. At the same time, none of the automotive interviewees have direct access to or actively aggregate field user interaction data. These circumstances also show automotive UX experts mainly rely on explicit feedback from users and do not leverage implicit feedback through field user interaction data. As mentioned before, with the tracking of user interaction data on company vehicles, interviewee P2 reports the most advanced data-driven processes of the automotive participants.

In the post-design phase, 7 out of 8 automotive participants confirmed they do not get consistent and detailed feedback based on field user interaction on their product. Five of these practitioners, however, confirm that they do get feedback through market analyses and general customer surveys.
Interviewee P4 describes the feedback process in the post-design stage as follows: \textit{``At the moment, we only receive feedback through studies [the company] executes, that take weeks and months. They take the [product], test it in multiple markets with many people, and curate a [report] with the results.''}
In the digital domain, 4 out of 6 practitioners have implemented a process to receive user feedback for their products based on field user interaction data.

\begin{resp}{\textbf{Key findings:\\}}
	\textbf{State-of-the-art in automotive (RQ1):} 
	\begin{itemize}
		\item Requirements and feature elicitation is not user-centered and only supported by general market research.
		\item Focus is on explicit feedback; implicit feedback through field user interaction data is not evident.
		\item Prototypes are mainly evaluated qualitatively by co-workers and in-house UX experts.
	\end{itemize}
	\tcblower
	{\textbf{Differences to digital companies (RQ4):}} 
	\begin{itemize}
		\item In digital companies, field user interaction data is considered for decisions throughout all design phases.
		\item For feature elicitation and prioritization, digital companies use a mixture of explicit feedback from representative users and implicit insights from field user interaction data.
		\item Within digital companies, insights from field user interaction data are very broad and range from feature usages to sophisticated hypotheses testing.
	\end{itemize}
\end{resp}

\subsection{Challenges}

\Cref{fig:BarPlot} shows that \textbf{Management Opinions} are often considered a challenge in the automotive domain. In particular, in the pre-design phase, six out of eight participants from the automotive domain report that their findings from user research are not considered in the decision making process. They argue that their proposals are often overruled by higher management even though they provide evidence through their research. Practitioners from digital companies do not experience this challenge as often. This correlates with the challenge to back up the designer's hypotheses toward user interaction with the product (\textbf{Backup of Hypothesis}). This challenge was expressed only by automotive participants. P7 states: \textit{``There are a lot of assumptions that people make about who is driving our cars, but none of them is actually backed up with any kind of information''}.
\textbf{Data Accessibility} and \textbf{Data Collection} are also mentioned frequently in the pre-design phase. Data accessibility refers to a generally insufficient availability, i.e. accessibility of user-related data within the company. P8 mentions that \textit{``[t]here is a very strong silo mentality in companies in the acquisition of information, but also in the distribution of information''}. The fact that all citations tagged with \textbf{Data Accessibility} come from automotive participants highlights the significant deficits regarding data transparency.
Considering the data collection challenges, all participants mainly refer to the challenge of collecting data as detailed as possible without violating legal restrictions. However, there are further technical peculiarities that complicate extensive data collection from users in the automotive domain.
P7 states that for \textit{``the older systems none of this existed, so we have no way of understanding what people did with it''}. Additionally challenging is the need for long-lasting architectures and the heterogeneous data processed by multiple different \acp{ECU}. P2 exemplifies that the current architecture of their vehicle platform is not yet prepared for the kind of interaction logging needed today.

\textbf{Time Restriction} and \textbf{Design Decisions} are the most often mentioned challenges by all participants regarding the design phase. Six participants describe that they often lack time to dive deep into user studies or interaction data. P1 explains: \textit{``The first priority is speed. We can't work on data for two or three days''}. Considering design decisions, 5 out of 8 automotive participants see a significant challenge in evaluating their designs and prototypes with regard to the context, i.e. the driving situation (\textbf{Context Consideration}). The participants further describe that the driving task itself and the influences from the environment affect how the driver interacts with the system. The difficulty of recreating this driving situation in a lab experiment is explained by P1: \textit{``The difference lies in the dual-task paradigm. When you are in the lab, the interaction with the HMI is the primary task, when you are driving it is only the secondary task''}. P5 adds that dynamic driving simulators offer the possibility to model the driving situation to a certain degree but that due to high cost and low availability they are only used for very few studies. The participants from digital companies focus on \textbf{Data Interpretability} and what methods need to be applied to make reliable statements.

In the post-design phase, the challenge of \textbf{Data Accessibility} reoccurs. Three automotive participants argue that the biggest challenge after a product's release is to get field user interaction data to evaluate how the product is accepted by the users. P2 states that \textit{``one of the main challenges is to make the right data available at the right time''}. In addition to the data being available, the challenges of intuitive \textbf{Data Presentation} are discussed by the participants as well. Six participants express that, due to the amount of data, field user interaction data needs to be visualized in an intuitively understandable way. P4 underlines this challenge by saying that \textit{``an 80-page pdf with results [...] doesn't help that much because nobody wants to read through it and it doesn't motivate designers to change anything''}.

\begin{resp}{\textbf{Key findings:\\}}
	\textbf{Challenges in automotive (RQ2):} 
	\begin{itemize}
		\item Field user interaction data is often not available or accessible throughout the design process due to organizational, legal, or technical restrictions.
		\item User research is not valued; Evidence-based design decisions are overruled by management.
		\item The complexity of the driving context further affects the already challenging task to create insights from large amounts of field user interaction data.
	\end{itemize}
	\tcblower
	{\textbf{Differences to digital companies (RQ4):}}
	\begin{itemize}
		\item The disparity between user insights and management opinions is less challenging for digital companies.
		\item Digital companies face more mature challenges in terms of integrating data in their design process rather than technical or organizational challenges.
	\end{itemize}
\end{resp}

\subsection{Needs}

The distribution of codes addressing the needs of the UX experts does not show major differences between the automotive and the digital domain (see \Cref{fig:BarPlot}). Most mentioned for both groups are explicit demands for data-driven \textbf{Feature}s. The features are manifold and range from dashboards visualizing feature-specific clickstreams to the implementation of data-driven analyses in design tools.

In the pre-design phase, 5 out of 8 participants from the automotive domain mention the explicit need for data-driven solutions to support their hypotheses and proposals made in early phases of development (\textbf{Empowerment}). This is connected to the state-of-the-art and the resulting challenges, that personal opinions in higher management play an important role in feature elicitation and prioritization. However, participants from digital companies do not express this need in the pre-design phase. They rather emphasize the significance of \textbf{Qualitative Feedback} and the need for \textbf{Triangulation} of different data sources. The need for qualitative data is important in early ideation phases, especially for new products. P10 states: \textit{ ``For a comprehensive redesign of a product you can't test A/B, you have to [\ldots] test them qualitatively to see if it makes sense to implement the hypothesis''}.

Revisiting the challenges of time restrictions and decision-making, 5 out of 8 automotive participants express a need to automatically evaluate their designs based on data retrieved from field usage (\textbf{Feature} and \textbf{Evaluation Metrics}). P9 agrees that such a feature would facilitate their advances toward a user-centered design approach: \textit{``[I]t can really help to defend my decisions. I guess, honestly, I'm always trying to defend it, not for myself but for the user''}. However, regarding automated analyses and models based on field user interaction data, especially the digital domain participants express a need for explainability. P10: \textit{``When you have some kind of magic box where I present a prototype and a magic score falls out, of course, people who are not so much into UX would ask: 'ok, but what does the box do? How does it get that number? Can I even trust it?' ''}.

The needs expressed most often in the post-design phase address how to measure the acceptance of a developed product or feature by field users. The participants indicate a need for \textbf{Evaluation Metric}s that quantify user acceptance and how it changes over time. Among conventional metrics like the number of clicks or conversion rates, 3 out of 6 participants from digital companies say that it is necessary to correlate these ratings with other \acp{KPI} like profit or newsletter subscriptions.

\begin{resp_float}{\textbf{Key findings:\\}}
	\textbf{Needs of automotive (RQ2)}: 
	\begin{itemize}
		\item Statistical support based on field user interaction data to leverage design hypotheses, feature elicitation, and prioritization.
		\item Tool support to automatically evaluate designs.
		\item Automated methods should offer explanations to facilitate interpretability.
	\end{itemize}
	\tcblower
	{\textbf{Differences to digital companies (RQ4):}} 
	\begin{itemize}
		\item In digital companies, many of the needs toward hypothesis support, feature elicitation, and feature acceptance assessment are already satisfied.
		\item In digital companies, there is a greater need to triangulate qualitative and quantitative data.
	\end{itemize}
\end{resp_float}

\subsection{Potentials}
In the pre-design phase, the automotive participants are particularly interested in the potential of \textbf{Feature Activation Analysis}, i.e. the evaluation of usage frequencies and duration. Especially for arguing against management opinions, 6 out of 8 automotive participants made statements that those analyses can satisfy the expressed need to empower them in their decisions. They explain that feature activation analyses accompanied by appropriate metrics can offer valuable insights into the field usage of features. Therefore, they can facilitate feature elicitation and prioritization. The participants further indicate that \textbf{User Flow Analysis} based on field user interaction data can provide a deeper understanding of how the users behave in the current system. P6 states: \textit{``{[W]e} are very good at building solutions but not always good at identifying the right problems''} and formulates the idea to \textit{``take the personas themselves from the market research and enrich them with certain usage data that are important to understand the user journey''}.

To overcome challenges in the design phase, 11 out of 14 participants indicate that automated design evaluation methods based on field user interaction data could offer valuable \textbf{Design Support}. This design support could be manifested in automated \fix{quantitative} usability analyses or the extraction of usage patterns for different user groups from extensive field data. 13 out of 14 participants indicate that the usage of field user interaction data for \textbf{User Modeling} could play an important role in their design process. P10 suggests using a \textit{``model that represents a persona to automatically evaluate a prototype''}. Another recurring theme is the topic of context consideration. 7 out of 8 automotive participants see the potential to use field user interaction data to analyze how the driving context affects user interactions with the product (\textbf{Context Analysis}). The interviewees argue that the context plays an important role in the automotive domain since the interaction with the environment is bidirectional. P1 states that it would be necessary to not only evaluate a feature based on its interaction data but also on how its usage influence the driving behavior. The latter has a direct and potentially fatal impact on its environment. This critical correlation could be evaluated by matching user interactions with driving data like lane-keeping parameters.

In the post-design phase, the participants see the biggest potential of field user interaction data in monitoring how features and products are accepted in the field. They argue that instant monitoring after release and an easy to understand data presentation would offer interesting insights into how often features are used and how the interaction changes over time. P1 elaborates on the direct connection to the subsequent pre-design phase: \textit{``Requirement analysis would also mean looking at the data that was collected at the end of the last version again. This should ideally be a cycle and I see the methods data-driven analyses offer at every point in this development lifecycle''}. 

\begin{resp_float}{\textbf{Key findings:\\}}
	\textbf{Potentials in automotive (RQ3):} 
	\begin{itemize}
		\item Insights from field user interaction data can shift the elicitation and prioritization of features from personal best guesses to \fix{more} user-centered decisions.
		\item Automated evaluation methods and user modeling based on field user interaction data may offer valuable design support.
		\item Field user interaction data can be triangulated with contextual data to investigate the impact of the driving situation on the interaction and vice versa, making evaluations less biased.
	\end{itemize}
		\tcblower
	{\textbf{Differences to digital companies (RQ4):}} 
	\begin{itemize}
		\item Most identified potentials apply to both, automotive and digital domains, but digital companies are more advanced in unlocking these potentials.
	\end{itemize}
\end{resp_float}

\subsection{Concerns}
In the pre-design phase, the participants express few concerns toward data-driven methods and the analysis of field user interaction data. P1 and P7 do not see any benefit of the discussed methods when it comes to the early ideation phases of a product. P7 states that \textit{``[t]hat's an interesting insight that maybe all the data-driven stuff has a bigger impact on everything where you try to optimize something in contrast to the work where the creative process is the main part''}.

The predominant concern in the design phase regards the \textbf{Limitation of Creativity} of the designers which might be caused by extensive use of data-driven analyses. This is strongly connected to the concerns in the pre-design stage, as participants from both groups see a risk to get stuck in small, iterative optimization processes. They anticipate that optimizing features based on historical data prevents thinking outside of the box to create something new. These concerns are related to the concerns toward \textbf{User Models}. P10 states that it is difficult to build a model without limiting creativity and describes it as an \textit{``overfitting problem: the model has seen too much old data and is therefore not able to generalize when it is applied to something new''}. Further, 7 participants (4 automotive) are concerned about how to interpret the results produced by an automated evaluation method (\textbf{Interpretability}). They mainly argue that an explanation has to be provided to develop trust in automatic evaluation \textit{``because usage scores alone produce very little insight''}, according to P12. P4 agrees by indicating: \textit{``A score might be ok, but there should be suggestions or information on how the score is calculated and influenced''}. 

There are very few concerns regarding the post-design phase. However, 5 out of 8 automotive participants communicate general doubts that legislation may prevent certain features and functions from being realized due to data collection restrictions. This especially holds for potentially person-related data, e.g. GPS coordinates of a vehicle.

\begin{resp_float}{\textbf{Key findings:\\}}
	\textbf{Concerns in automotive (RQ2):} 
	\begin{itemize}
	  \item Insurmountable organizational, legal, or technical restrictions prevent that data can be collected.
	  \item Quantitative insights may not be useful in early ideation phases to evaluate volatile concepts.
	  \item Data-driven methods may limit creativity and shift the focus to small incremental changes.
	\end{itemize}
		\tcblower
	{\textbf{Differences to digital companies (RQ4):}} 
	\begin{itemize}
	  \item Participants from digital companies expressed more concerns.
	  \item A lack of interpretability can lead to disuse of data-driven methods.
	\end{itemize}
\end{resp_float}

\section{Discussion}

In this section, we reflect on the needs, challenges, and concerns expressed by the practitioners and emphasize untapped potential in the evaluation and development of \acp{IVIS}. We additionally relate our findings to prior published research and present methods that may benefit the automotive UX development lifecycle.

Leadership, culture, and the mindset within a company highly influence the usage of data-driven methods. Compared to digital-native companies, automotive \acp{OEM} find it difficult to keep up when it comes to the integration of data-driven methodologies in UX development. However, data-driven methods based on field user interaction data can benefit the development and evaluation of \acp{IVIS}. In the pre-design phase, we see great potential in generating a deeper understanding of the users and their behavior through analyses of field user interaction data. Data-driven methods can facilitate decisions in early phases to prioritize features or products. Multiple approaches~\cite{Liu.2017, Wang.2016, Benevenuto.2009} that enhance the user understanding based on analyses of automatically collected field user interaction data can be leveraged to unlock this potential for \acp{IVIS}. \fix{However, as participants also mentioned, these approaches should be considered as an additional source of user feedback and not as a replacement for already existing qualitative methods.}

In the design phase, automated usability tests~\cite{Speicher.2015, Deka.2017, Ma.2012} can play an important role in making the design process more user-centered and efficient at the same time. The fact that the context of use, i.e. the driving situation, is inherently contained in field data is another key advantage. Additionally, the possibility to explicitly map field user interaction data with naturalistic driving data creates new opportunities in the design and evaluation of \acp{IVIS}. One can, for example, predict driver distraction~\cite{Salis.2019}, secondary task engagement~\cite{Risteska.2018}, or identify drivers based on their driving behavior~\cite{Zhang.2016}. This allows considering the complex interactions between driver, car, and environment without the costs and bias introduced by simulator experiments. This is in line with the findings made in earlier work on this topic~\cite{Orlovska.2018, Orlovska.2018b}. However, to provide the biggest possible value for practitioners, all automated methods should provide an explanatory component \fix{and be triangulated with qualitative user feedback}.

In the post-design phase, there is a need to monitor the acceptance and usage of \acp{IVIS} after deployment. Here, data-driven methods offer insights that can then benefit the next development cycle. 

However, in line with the findings of \citeauthor{Lamm.2019}~\cite{Lamm.2019}, automated and model-based approaches currently do not play an important role in the evaluation of \acp{IVIS}. This originates from none of the interviewed \acp{OEM} having a system in place that is explicitly developed to record detailed user interactions. Current systems are yet built for different purposes and only modified to offer basic capabilities, while dedicated systems are only available for test fleets. Legacy car architectures and long product life cycles aggravate the difficulty to implement such new systems. Additionally, strong restrictions regarding privacy and security are challenging for \acp{OEM}. However, a dedicated system for interaction logging that provides detailed and high-quality data is the cornerstone of the potentials introduced by field data-driven methods.

\section{Conclusion}

Our results show that data-driven methods based on field user interaction data can have great value for the automotive UX development lifecycle and can play an essential role in making the development of \acp{IVIS} more user-centered. All automotive domain experts in our study agree that there is a lack of implicit feedback through field user interaction data in UX development. These findings coincide with the work of \citeauthor{Orlovska.2018}~\cite{Orlovska.2018, Orlovska.2018b}. Additionally, our results support the disparity presented in \Cref{subsec:rel_work} that in comparison to the automotive domain, digital domains are far ahead in exploring the potentials of field user interaction data. \fix{We conclude that the transition from predominately explicit and qualitative user feedback, e.g. through customer surveys or studies, to a combination of the former with implicit feedback through automatically collected field user interaction data is a necessary process for the development of an automotive UX fulfilling users' diverse needs.} Another important benefit is the statistical support of designers' decisions to overcome the current opinionated guesswork often encountered in automotive UX development. Interestingly, in the automotive and the digital domain, we identified a high potential for the automated evaluation of field user interaction data and advanced user modeling based on interaction data for early prototype evaluation. These identified potentials will be subject to future work to unlock the benefits field user interaction data offers. Finally, the results of this work facilitate research on data-driven methods in the automotive UX development by directing the focus toward the needs, challenges, and concerns UX experts face today.

\bibliographystyle{ACM-Reference-Format}
\bibliography{UXStudy}


\begin{thebibliography}{49}


\ifx \showCODEN    \undefined \def \showCODEN     #1{\unskip}     \fi
\ifx \showDOI      \undefined \def \showDOI       #1{#1}\fi
\ifx \showISBNx    \undefined \def \showISBNx     #1{\unskip}     \fi
\ifx \showISBNxiii \undefined \def \showISBNxiii  #1{\unskip}     \fi
\ifx \showISSN     \undefined \def \showISSN      #1{\unskip}     \fi
\ifx \showLCCN     \undefined \def \showLCCN      #1{\unskip}     \fi
\ifx \shownote     \undefined \def \shownote      #1{#1}          \fi
\ifx \showarticletitle \undefined \def \showarticletitle #1{#1}   \fi
\ifx \showURL      \undefined \def \showURL       {\relax}        \fi
\providecommand\bibfield[2]{#2}
\providecommand\bibinfo[2]{#2}
\providecommand\natexlab[1]{#1}
\providecommand\showeprint[2][]{arXiv:#2}

\bibitem[\protect\citeauthoryear{Ahmad, Langdon, Godsill, Hardy, Skrypchuk, and
  Donkor}{Ahmad et~al\mbox{.}}{2015}]%
        {Ahmad.2015}
\bibfield{author}{\bibinfo{person}{Bashar~I. Ahmad},
  \bibinfo{person}{Patrick~M. Langdon}, \bibinfo{person}{Simon~J. Godsill},
  \bibinfo{person}{Robert Hardy}, \bibinfo{person}{Lee Skrypchuk}, {and}
  \bibinfo{person}{Richard Donkor}.} \bibinfo{year}{2015}\natexlab{}.
\newblock \showarticletitle{Touchscreen usability and input performance in
  vehicles under different road conditions}. In
  \bibinfo{booktitle}{\emph{Proceedings of the 7th International Conference on
  Automotive User Interfaces and Interactive Vehicular Applications -
  {AutomotiveUI} {\textquotesingle}15}}. \bibinfo{publisher}{{ACM} Press},
  \bibinfo{address}{Nottingham}, \bibinfo{pages}{47--50}.
\newblock
\urldef\tempurl%
\url{https://doi.org/10.1145/2799250.2799284}
\showDOI{\tempurl}


\bibitem[\protect\citeauthoryear{Alvarez, Jordan, Knopf, LeBlanc, Rumbel, and
  Zafiroglu}{Alvarez et~al\mbox{.}}{2017}]%
        {Alvarez.2017}
\bibfield{author}{\bibinfo{person}{Ignacio Alvarez}, \bibinfo{person}{Adam
  Jordan}, \bibinfo{person}{Juliana Knopf}, \bibinfo{person}{Darrell LeBlanc},
  \bibinfo{person}{Laura Rumbel}, {and} \bibinfo{person}{Alexandra Zafiroglu}.}
  \bibinfo{year}{2017}\natexlab{}.
\newblock \showarticletitle{The
  Insight{\textendash}Prototype{\textendash}Product Cycle Best Practices and
  Processes to Iteratively Advance In-Vehicle Interactive Experiences
  Development}.
\newblock In \bibinfo{booktitle}{\emph{Automotive User Interfaces}}.
  \bibinfo{publisher}{Springer International Publishing},
  \bibinfo{address}{Cham, Switzerland}, \bibinfo{pages}{377--400}.
\newblock
\urldef\tempurl%
\url{https://doi.org/10.1007/978-3-319-49448-7_14}
\showDOI{\tempurl}


\bibitem[\protect\citeauthoryear{Atterer, Wnuk, and Schmidt}{Atterer
  et~al\mbox{.}}{2006}]%
        {Atterer.2006}
\bibfield{author}{\bibinfo{person}{Richard Atterer}, \bibinfo{person}{Monika
  Wnuk}, {and} \bibinfo{person}{Albrecht Schmidt}.}
  \bibinfo{year}{2006}\natexlab{}.
\newblock \showarticletitle{Knowing the user{\textquotesingle}s every move}. In
  \bibinfo{booktitle}{\emph{Proceedings of the 15th international conference on
  World Wide Web - {WWW} {\textquotesingle}06}}. \bibinfo{publisher}{{ACM}
  Press}, \bibinfo{address}{Edinburgh, Scotland}, \bibinfo{pages}{203--212}.
\newblock
\urldef\tempurl%
\url{https://doi.org/10.1145/1135777.1135811}
\showDOI{\tempurl}


\bibitem[\protect\citeauthoryear{Baker, Crusius, Fischer, Gerling,
  Gnanaserakan, Kerstan, Kuhnert, Kusber, Mohs, Schulte, Seyfferth, Stephan,
  and Warnke}{Baker et~al\mbox{.}}{2016}]%
        {Baker.2016}
\bibfield{author}{\bibinfo{person}{Edward~H Baker}, \bibinfo{person}{David
  Crusius}, \bibinfo{person}{Marco Fischer}, \bibinfo{person}{Walter Gerling},
  \bibinfo{person}{Kaushik Gnanaserakan}, \bibinfo{person}{Henning Kerstan},
  \bibinfo{person}{Felix Kuhnert}, \bibinfo{person}{Julia Kusber},
  \bibinfo{person}{Joachim Mohs}, \bibinfo{person}{Manuel Schulte},
  \bibinfo{person}{Jonas Seyfferth}, \bibinfo{person}{Juliane Stephan}, {and}
  \bibinfo{person}{Trent Warnke}.} \bibinfo{year}{2016}\natexlab{}.
\newblock \bibinfo{title}{{Connected car report 2016: Opportunities, risk, and
  turmoil on the road to autonomous vehicles}}.
\newblock
\newblock
\urldef\tempurl%
\url{https://www.strategyand.pwc.com/gx/en/insights/2016/connected-car-2016-study.html}
\showURL{%
\tempurl}


\bibitem[\protect\citeauthoryear{Benevenuto, Rodrigues, Cha, and
  Almeida}{Benevenuto et~al\mbox{.}}{2009}]%
        {Benevenuto.2009}
\bibfield{author}{\bibinfo{person}{Fabr{\'i}cio Benevenuto},
  \bibinfo{person}{Tiago Rodrigues}, \bibinfo{person}{Meeyoung Cha}, {and}
  \bibinfo{person}{Virg{\'i}lio Almeida}.} \bibinfo{year}{2009}\natexlab{}.
\newblock \showarticletitle{{Characterizing user behavior in online social
  networks}}. In \bibinfo{booktitle}{\emph{{Proceedings of the 9th ACM SIGCOMM
  conference on Internet measurement conference}}},
  \bibfield{editor}{\bibinfo{person}{Anja Feldmann}} (Ed.).
  \bibinfo{publisher}{ACM}, \bibinfo{address}{New York, NY},
  \bibinfo{pages}{49}.
\newblock
\showISBNx{9781605587714}
\urldef\tempurl%
\url{https://doi.org/10.1145/1644893.1644900}
\showDOI{\tempurl}


\bibitem[\protect\citeauthoryear{Broy}{Broy}{2006}]%
        {Broy.2006}
\bibfield{author}{\bibinfo{person}{Manfred Broy}.}
  \bibinfo{year}{2006}\natexlab{}.
\newblock \showarticletitle{{Challenges in Automotive Software Engineering}}.
  In \bibinfo{booktitle}{\emph{{Proceedings of the 28th International
  Conference on Software Engineering}}} \emph{(\bibinfo{series}{{ICSE '06}})}.
  \bibinfo{publisher}{{Association for Computing Machinery}},
  \bibinfo{address}{New York, NY, USA}, \bibinfo{pages}{33--42}.
\newblock
\showISBNx{1595933751}
\urldef\tempurl%
\url{https://doi.org/10.1145/1134285.1134292}
\showDOI{\tempurl}


\bibitem[\protect\citeauthoryear{de~Salis, Baumgartner, and Carrino}{de~Salis
  et~al\mbox{.}}{2019}]%
        {Salis.2019}
\bibfield{author}{\bibinfo{person}{Emmanuel de Salis},
  \bibinfo{person}{Dan~Yvan Baumgartner}, {and} \bibinfo{person}{Stefano
  Carrino}.} \bibinfo{year}{2019}\natexlab{}.
\newblock \showarticletitle{Can we predict driver distraction without driver
  psychophysiological state?}. In \bibinfo{booktitle}{\emph{Proceedings of the
  11th International Conference on Automotive User Interfaces and Interactive
  Vehicular Applications Adjunct Proceedings - {AutomotiveUI}
  {\textquotesingle}19}}. \bibinfo{publisher}{{ACM} Press},
  \bibinfo{address}{Utrecht Netherlands}, \bibinfo{pages}{194--198}.
\newblock
\urldef\tempurl%
\url{https://doi.org/10.1145/3349263.3351514}
\showDOI{\tempurl}


\bibitem[\protect\citeauthoryear{Deka, Huang, Franzen, Nichols, Li, and
  Kumar}{Deka et~al\mbox{.}}{2017}]%
        {Deka.2017}
\bibfield{author}{\bibinfo{person}{Biplab Deka}, \bibinfo{person}{Zifeng
  Huang}, \bibinfo{person}{Chad Franzen}, \bibinfo{person}{Jeffrey Nichols},
  \bibinfo{person}{Yang Li}, {and} \bibinfo{person}{Ranjitha Kumar}.}
  \bibinfo{year}{2017}\natexlab{}.
\newblock \showarticletitle{{ZIPT}}. In \bibinfo{booktitle}{\emph{Proceedings
  of the 30th Annual {ACM} Symposium on User Interface Software and Technology
  - {UIST} {\textquotesingle}17}}. \bibinfo{publisher}{{ACM} Press},
  \bibinfo{address}{Quebec City QC Canada}, \bibinfo{pages}{727--736}.
\newblock
\urldef\tempurl%
\url{https://doi.org/10.1145/3126594.3126647}
\showDOI{\tempurl}


\bibitem[\protect\citeauthoryear{Ebel, G{\"o}l, Lingenfelder, and
  Vogelsang}{Ebel et~al\mbox{.}}{2020}]%
        {Ebel.2020}
\bibfield{author}{\bibinfo{person}{Patrick Ebel}, \bibinfo{person}{Ibrahim~Emre
  G{\"o}l}, \bibinfo{person}{Christoph Lingenfelder}, {and}
  \bibinfo{person}{Andreas Vogelsang}.} \bibinfo{year}{2020}\natexlab{}.
\newblock \showarticletitle{Destination Prediction Based on Partial Trajectory
  Data}. In \bibinfo{booktitle}{\emph{\emph{accepted at:} IEEE Intelligent
  Vehicles Symposium (IV)}}. \bibinfo{publisher}{arXiv.org},
  \bibinfo{address}{Las Vegas Nevada}, \bibinfo{pages}{1--7}.
\newblock
\urldef\tempurl%
\url{https://arxiv.org/abs/2004.07473}
\showURL{%
\tempurl}


\bibitem[\protect\citeauthoryear{Etikan}{Etikan}{2016}]%
        {Etikan.2016}
\bibfield{author}{\bibinfo{person}{Ilker Etikan}.}
  \bibinfo{year}{2016}\natexlab{}.
\newblock \showarticletitle{Comparison of convenience sampling and purposive
  sampling}.
\newblock \bibinfo{journal}{\emph{American journal of theoretical and applied
  statistics}} \bibinfo{volume}{5}, \bibinfo{number}{1} (\bibinfo{year}{2016}),
  \bibinfo{pages}{1--4}.
\newblock


\bibitem[\protect\citeauthoryear{Fastrez and Hau{\'{e}}}{Fastrez and
  Hau{\'{e}}}{2008}]%
        {Fastrez.2008}
\bibfield{author}{\bibinfo{person}{Pierre Fastrez} {and}
  \bibinfo{person}{Jean-Baptiste Hau{\'{e}}}.} \bibinfo{year}{2008}\natexlab{}.
\newblock \showarticletitle{Designing and evaluating driver support systems
  with the user in mind}.
\newblock \bibinfo{journal}{\emph{International Journal of Human-Computer
  Studies}} \bibinfo{volume}{66}, \bibinfo{number}{3} (\bibinfo{date}{mar}
  \bibinfo{year}{2008}), \bibinfo{pages}{125--131}.
\newblock
\urldef\tempurl%
\url{https://doi.org/10.1016/j.ijhcs.2008.01.001}
\showDOI{\tempurl}


\bibitem[\protect\citeauthoryear{Frison, Wintersberger, and Riener}{Frison
  et~al\mbox{.}}{2019a}]%
        {Frison.2019}
\bibfield{author}{\bibinfo{person}{Anna~Katharina Frison},
  \bibinfo{person}{Philipp Wintersberger}, {and} \bibinfo{person}{Andreas
  Riener}.} \bibinfo{year}{2019}\natexlab{a}.
\newblock \showarticletitle{{Resurrecting the ghost in the shell: A
  need-centered development approach for optimizing user experience in highly
  automated vehicles}}.
\newblock \bibinfo{journal}{\emph{{Transportation Research Part F Traffic
  Psychology and Behaviour}}}  \bibinfo{volume}{65} (\bibinfo{year}{2019}),
  \bibinfo{pages}{439--456}.
\newblock
\urldef\tempurl%
\url{https://doi.org/10.1016/j.trf.2019.08.001}
\showDOI{\tempurl}


\bibitem[\protect\citeauthoryear{Frison, Wintersberger, Riener,
  Schartm{\"u}ller, Boyle, Miller, and Weigl}{Frison et~al\mbox{.}}{2019b}]%
        {Frison.2019b}
\bibfield{author}{\bibinfo{person}{Anna-Katharina Frison},
  \bibinfo{person}{Philipp Wintersberger}, \bibinfo{person}{Andreas Riener},
  \bibinfo{person}{Clemens Schartm{\"u}ller}, \bibinfo{person}{Linda~Ng Boyle},
  \bibinfo{person}{Erika Miller}, {and} \bibinfo{person}{Klemens Weigl}.}
  \bibinfo{year}{2019}\natexlab{b}.
\newblock \showarticletitle{{In UX We Trust}}. In
  \bibinfo{booktitle}{\emph{{Proceedings of the 2019 CHI Conference on Human
  Factors in Computing Systems - CHI '19}}},
  \bibfield{editor}{\bibinfo{person}{Stephen Brewster},
  \bibinfo{person}{Geraldine Fitzpatrick}, \bibinfo{person}{Anna Cox}, {and}
  \bibinfo{person}{Vassilis Kostakos}} (Eds.). \bibinfo{publisher}{{ACM
  Press}}, \bibinfo{address}{New York, New York, USA}, \bibinfo{pages}{1--13}.
\newblock
\showISBNx{9781450359702}
\urldef\tempurl%
\url{https://doi.org/10.1145/3290605.3300374}
\showDOI{\tempurl}


\bibitem[\protect\citeauthoryear{Gerber, Schroeter, and Vehns}{Gerber
  et~al\mbox{.}}{2019}]%
        {Gerber.2019}
\bibfield{author}{\bibinfo{person}{Michael~A. Gerber}, \bibinfo{person}{Ronald
  Schroeter}, {and} \bibinfo{person}{Julia Vehns}.}
  \bibinfo{year}{2019}\natexlab{}.
\newblock \showarticletitle{A Video-Based Automated Driving Simulator for
  Automotive {UI} Prototyping, {UX} and Behaviour Research}. In
  \bibinfo{booktitle}{\emph{Proceedings of the 11th International Conference on
  Automotive User Interfaces and Interactive Vehicular Applications -
  {AutomotiveUI} {\textquotesingle}19}}. \bibinfo{publisher}{{ACM} Press},
  \bibinfo{address}{Utrecht Netherlands}, \bibinfo{pages}{14--23}.
\newblock
\urldef\tempurl%
\url{https://doi.org/10.1145/3342197.3344533}
\showDOI{\tempurl}


\bibitem[\protect\citeauthoryear{Gray}{Gray}{2014}]%
        {Gray.2014}
\bibfield{author}{\bibinfo{person}{Colin~M. Gray}.}
  \bibinfo{year}{2014}\natexlab{}.
\newblock \showarticletitle{Evolution of design competence in {UX} practice}.
  In \bibinfo{booktitle}{\emph{Proceedings of the 32nd annual {ACM} conference
  on Human factors in computing systems - {CHI} {\textquotesingle}14}}.
  \bibinfo{publisher}{{ACM} Press}, \bibinfo{address}{Toronto Ontario Canada},
  \bibinfo{pages}{1645--1654}.
\newblock
\urldef\tempurl%
\url{https://doi.org/10.1145/2556288.2557264}
\showDOI{\tempurl}


\bibitem[\protect\citeauthoryear{Guest, Bunce, and Johnson}{Guest
  et~al\mbox{.}}{2006}]%
        {Guest.2006}
\bibfield{author}{\bibinfo{person}{Greg Guest}, \bibinfo{person}{Arwen Bunce},
  {and} \bibinfo{person}{Laura Johnson}.} \bibinfo{year}{2006}\natexlab{}.
\newblock \showarticletitle{How Many Interviews Are Enough?: An Experiment with
  Data Saturation and Variability}.
\newblock \bibinfo{journal}{\emph{Field Methods}} \bibinfo{volume}{18},
  \bibinfo{number}{1} (\bibinfo{date}{feb} \bibinfo{year}{2006}),
  \bibinfo{pages}{59--82}.
\newblock
\urldef\tempurl%
\url{https://doi.org/10.1177/1525822x05279903}
\showDOI{\tempurl}


\bibitem[\protect\citeauthoryear{Harvey and Stanton}{Harvey and
  Stanton}{2016}]%
        {Harvey.2016}
\bibfield{author}{\bibinfo{person}{Catherine Harvey} {and}
  \bibinfo{person}{Neville~A. Stanton}.} \bibinfo{year}{2016}\natexlab{}.
\newblock \bibinfo{booktitle}{\emph{Usability Evaluation for In-Vehicle
  Systems}}.
\newblock \bibinfo{publisher}{Taylor \& Francis Ltd.}, \bibinfo{address}{Boca
  Raton Florida}.
\newblock
\showISBNx{1466514302}
\urldef\tempurl%
\url{https://www.ebook.de/de/product/21182820/catherine_harvey_neville_a_stanton_usability_evaluation_for_in_vehicle_systems.html}
\showURL{%
\tempurl}


\bibitem[\protect\citeauthoryear{Harvey, Stanton, Pickering, McDonald, and
  Zheng}{Harvey et~al\mbox{.}}{2010}]%
        {Harvey.2010}
\bibfield{author}{\bibinfo{person}{Catherine Harvey},
  \bibinfo{person}{Neville~A. Stanton}, \bibinfo{person}{Carl~A. Pickering},
  \bibinfo{person}{Michael McDonald}, {and} \bibinfo{person}{Pengjun Zheng}.}
  \bibinfo{year}{2010}\natexlab{}.
\newblock \showarticletitle{Context of use as a factor in determining the
  usability of in-vehicle devices}.
\newblock \bibinfo{journal}{\emph{Theoretical Issues in Ergonomics Science}}
  \bibinfo{volume}{12}, \bibinfo{number}{4} (\bibinfo{date}{jun}
  \bibinfo{year}{2010}), \bibinfo{pages}{318--338}.
\newblock
\urldef\tempurl%
\url{https://doi.org/10.1080/14639221003717024}
\showDOI{\tempurl}


\bibitem[\protect\citeauthoryear{Harvey, Stanton, Pickering, McDonald, and
  Zheng}{Harvey et~al\mbox{.}}{2011}]%
        {Harvey.2011}
\bibfield{author}{\bibinfo{person}{Catherine Harvey},
  \bibinfo{person}{Neville~A. Stanton}, \bibinfo{person}{Carl~A. Pickering},
  \bibinfo{person}{Mike McDonald}, {and} \bibinfo{person}{Pengjun Zheng}.}
  \bibinfo{year}{2011}\natexlab{}.
\newblock \showarticletitle{A Usability Evaluation Toolkit for In-Vehicle
  Information Systems ({IVISs})}.
\newblock \bibinfo{journal}{\emph{Applied Ergonomics}} \bibinfo{volume}{42},
  \bibinfo{number}{4} (\bibinfo{date}{may} \bibinfo{year}{2011}),
  \bibinfo{pages}{563--574}.
\newblock
\urldef\tempurl%
\url{https://doi.org/10.1016/j.apergo.2010.09.013}
\showDOI{\tempurl}


\bibitem[\protect\citeauthoryear{Hassenzahl and Tractinsky}{Hassenzahl and
  Tractinsky}{2006}]%
        {Hassenzahl.2006}
\bibfield{author}{\bibinfo{person}{Marc Hassenzahl} {and} \bibinfo{person}{Noam
  Tractinsky}.} \bibinfo{year}{2006}\natexlab{}.
\newblock \showarticletitle{User experience~-~a research agenda}.
\newblock \bibinfo{journal}{\emph{Behaviour {\&} Information Technology}}
  \bibinfo{volume}{25}, \bibinfo{number}{2} (\bibinfo{date}{mar}
  \bibinfo{year}{2006}), \bibinfo{pages}{91--97}.
\newblock
\urldef\tempurl%
\url{https://doi.org/10.1080/01449290500330331}
\showDOI{\tempurl}


\bibitem[\protect\citeauthoryear{Heimgärtner, Solanki, and Windl}{Heimgärtner
  et~al\mbox{.}}{2017}]%
        {Heimgaertner.2017}
\bibfield{author}{\bibinfo{person}{Rüdiger Heimgärtner},
  \bibinfo{person}{Alkesh Solanki}, {and} \bibinfo{person}{Helmut Windl}.}
  \bibinfo{year}{2017}\natexlab{}.
\newblock \showarticletitle{Cultural User Experience in the
  Car{\textemdash}Toward a Standardized Systematic Intercultural Agile
  Automotive {UI}/{UX} Design Process}.
\newblock In \bibinfo{booktitle}{\emph{Automotive User Interfaces}}.
  \bibinfo{publisher}{Springer International Publishing},
  \bibinfo{address}{Cham Switzerland}, \bibinfo{pages}{143--184}.
\newblock
\urldef\tempurl%
\url{https://doi.org/10.1007/978-3-319-49448-7_6}
\showDOI{\tempurl}


\bibitem[\protect\citeauthoryear{K{\"o}rber and Bengler}{K{\"o}rber and
  Bengler}{2013}]%
        {Korber.2013b}
\bibfield{author}{\bibinfo{person}{Moritz K{\"o}rber} {and}
  \bibinfo{person}{Klaus Bengler}.} \bibinfo{year}{2013}\natexlab{}.
\newblock \showarticletitle{{Measurement of Momentary User Experience in an
  Automotive Context}}. In \bibinfo{booktitle}{\emph{{Proceedings of the 5th
  International Conference on Automotive User Interfaces and Interactive
  Vehicular Applications}}} \emph{(\bibinfo{series}{{AutomotiveUI '13}})}.
  \bibinfo{publisher}{{Association for Computing Machinery}},
  \bibinfo{address}{New York, NY, USA}, \bibinfo{pages}{194--201}.
\newblock
\showISBNx{9781450324786}
\urldef\tempurl%
\url{https://doi.org/10.1145/2516540.2516555}
\showDOI{\tempurl}


\bibitem[\protect\citeauthoryear{K{\"o}rber, Eichinger, Bengler, and
  Olaverri-Monreal}{K{\"o}rber et~al\mbox{.}}{2013}]%
        {Korber.2013}
\bibfield{author}{\bibinfo{person}{Moritz K{\"o}rber}, \bibinfo{person}{Armin
  Eichinger}, \bibinfo{person}{Klaus Bengler}, {and} \bibinfo{person}{Cristina
  Olaverri-Monreal}.} \bibinfo{year}{2013}\natexlab{}.
\newblock \showarticletitle{{User experience evaluation in an automotive
  context}}. In \bibinfo{booktitle}{\emph{{2013 IEEE Intelligent Vehicles
  Symposium Workshops (IV Workshops)}}}. \bibinfo{publisher}{Institute of
  Electrical and Electronics Engineers}, \bibinfo{address}{Gold Coast City,
  Australia}, \bibinfo{pages}{13--18}.
\newblock
\urldef\tempurl%
\url{https://doi.org/10.1109/IVWorkshops.2013.6615219}
\showDOI{\tempurl}


\bibitem[\protect\citeauthoryear{Krippendorff}{Krippendorff}{2018}]%
        {Krippendorff.2018}
\bibfield{author}{\bibinfo{person}{Klaus Krippendorff}.}
  \bibinfo{year}{2018}\natexlab{}.
\newblock \bibinfo{booktitle}{\emph{Content analysis: An introduction to its
  methodology}}.
\newblock \bibinfo{publisher}{Sage publications}, \bibinfo{address}{Thousand
  Oaks, California}.
\newblock


\bibitem[\protect\citeauthoryear{Lamm and Wolff}{Lamm and Wolff}{2019}]%
        {Lamm.2019}
\bibfield{author}{\bibinfo{person}{Lukas Lamm} {and} \bibinfo{person}{Christian
  Wolff}.} \bibinfo{year}{2019}\natexlab{}.
\newblock \showarticletitle{{Exploratory Analysis of the Research Literature on
  Evaluation of In-Vehicle Systems Interfaces and Interactive Vehicular
  Applications, AutomotiveUI 2019, Utrecht, The Netherlands, September 21-25,
  2019}}. In \bibinfo{booktitle}{\emph{{Proceedings of the 11th International
  Conference on Automotive User Interfaces and Interactive Vehicular
  Applications, AutomotiveUI 2019, Utrecht, The Netherlands, September 21-25,
  2019}}}, \bibfield{editor}{\bibinfo{person}{Christian~P. Janssen},
  \bibinfo{person}{Stella~F. Donker}, \bibinfo{person}{Lewis~L. Chuang}, {and}
  \bibinfo{person}{Wendy Ju}} (Eds.). \bibinfo{publisher}{ACM},
  \bibinfo{address}{Utrecht, Netherlands}, \bibinfo{pages}{60--69}.
\newblock
\showISBNx{978-1-4503-6884-1}
\urldef\tempurl%
\url{https://doi.org/10.1145/3342197.3344527}
\showDOI{\tempurl}


\bibitem[\protect\citeauthoryear{Law, Roto, Hassenzahl, Vermeeren, and
  Kort}{Law et~al\mbox{.}}{2009}]%
        {Law.2009}
\bibfield{author}{\bibinfo{person}{Lai-Chong Law}, \bibinfo{person}{Virpi
  Roto}, \bibinfo{person}{Marc Hassenzahl}, \bibinfo{person}{Arnold Vermeeren},
  {and} \bibinfo{person}{Joke Kort}.} \bibinfo{year}{2009}\natexlab{}.
\newblock \showarticletitle{Understanding, scoping and defining user
  experience: A survey approach}, In \bibinfo{booktitle}{Proceedings of the
  SIGCHI conference on human factors in computing systems}.
\newblock \bibinfo{journal}{\emph{Proc. CHI '09}} \bibinfo{volume}{1},
  \bibinfo{number}{27}, \bibinfo{pages}{719--728}.
\newblock
\urldef\tempurl%
\url{https://doi.org/10.1145/1518701.1518813}
\showDOI{\tempurl}


\bibitem[\protect\citeauthoryear{Lewis}{Lewis}{2009}]%
        {Lewis.2009}
\bibfield{author}{\bibinfo{person}{John Lewis}.}
  \bibinfo{year}{2009}\natexlab{}.
\newblock \showarticletitle{{Redefining Qualitative Methods: Believability in
  the Fifth Moment}}.
\newblock \bibinfo{journal}{\emph{{International Journal of Qualitative
  Methods}}} \bibinfo{volume}{8}, \bibinfo{number}{2} (\bibinfo{year}{2009}),
  \bibinfo{pages}{1--14}.
\newblock
\showISSN{1609-4069}
\urldef\tempurl%
\url{https://doi.org/10.1177/160940690900800201}
\showDOI{\tempurl}


\bibitem[\protect\citeauthoryear{Liu, Wang, Dontcheva, Hoffman, Walker, and
  Wilson}{Liu et~al\mbox{.}}{2017}]%
        {Liu.2017}
\bibfield{author}{\bibinfo{person}{Zhicheng Liu}, \bibinfo{person}{Yang Wang},
  \bibinfo{person}{Mira Dontcheva}, \bibinfo{person}{Matthew Hoffman},
  \bibinfo{person}{Seth Walker}, {and} \bibinfo{person}{Alan Wilson}.}
  \bibinfo{year}{2017}\natexlab{}.
\newblock \showarticletitle{{Patterns and Sequences: Interactive Exploration of
  Clickstreams to Understand Common Visitor Paths}}.
\newblock \bibinfo{journal}{\emph{{IEEE transactions on visualization and
  computer graphics}}} \bibinfo{volume}{23}, \bibinfo{number}{1}
  (\bibinfo{year}{2017}), \bibinfo{pages}{321--330}.
\newblock
\urldef\tempurl%
\url{https://doi.org/10.1109/TVCG.2016.2598797}
\showDOI{\tempurl}


\bibitem[\protect\citeauthoryear{Löcken, Borojeni, Müller, Gable, Triberti,
  Diels, Glatz, Alvarez, Chuang, and Boll}{Löcken et~al\mbox{.}}{2017}]%
        {Loecken.2017}
\bibfield{author}{\bibinfo{person}{Andreas Löcken},
  \bibinfo{person}{Shadan~Sadeghian Borojeni}, \bibinfo{person}{Heiko Müller},
  \bibinfo{person}{Thomas~M. Gable}, \bibinfo{person}{Stefano Triberti},
  \bibinfo{person}{Cyriel Diels}, \bibinfo{person}{Christiane Glatz},
  \bibinfo{person}{Ignacio Alvarez}, \bibinfo{person}{Lewis Chuang}, {and}
  \bibinfo{person}{Susanne Boll}.} \bibinfo{year}{2017}\natexlab{}.
\newblock \showarticletitle{Towards Adaptive Ambient In-Vehicle Displays and
  Interactions: Insights and Design Guidelines from the 2015 {AutomotiveUI}
  Dedicated Workshop}.
\newblock In \bibinfo{booktitle}{\emph{Automotive User Interfaces}}.
  \bibinfo{publisher}{Springer International Publishing},
  \bibinfo{address}{Nottingham, UK}, \bibinfo{pages}{325--348}.
\newblock
\urldef\tempurl%
\url{https://doi.org/10.1007/978-3-319-49448-7_12}
\showDOI{\tempurl}


\bibitem[\protect\citeauthoryear{Ma, Yan, Chen, Zhang, Huang, Drury, and
  Wang}{Ma et~al\mbox{.}}{2012}]%
        {Ma.2012}
\bibfield{author}{\bibinfo{person}{Xiaoxiao Ma}, \bibinfo{person}{Bo Yan},
  \bibinfo{person}{Guanling Chen}, \bibinfo{person}{Chunhui Zhang},
  \bibinfo{person}{Ke Huang}, \bibinfo{person}{Jill Drury}, {and}
  \bibinfo{person}{Linzhang Wang}.} \bibinfo{year}{2012}\natexlab{}.
\newblock \showarticletitle{Design and Implementation of a Toolkit for
  Usability Testing of Mobile Apps}.
\newblock \bibinfo{journal}{\emph{Mobile Networks and Applications}}
  \bibinfo{volume}{18}, \bibinfo{number}{1} (\bibinfo{date}{nov}
  \bibinfo{year}{2012}), \bibinfo{pages}{81--97}.
\newblock
\urldef\tempurl%
\url{https://doi.org/10.1007/s11036-012-0421-z}
\showDOI{\tempurl}


\bibitem[\protect\citeauthoryear{Maxwell}{Maxwell}{2012}]%
        {Maxwell.2012}
\bibfield{author}{\bibinfo{person}{Joseph~A Maxwell}.}
  \bibinfo{year}{2012}\natexlab{}.
\newblock \bibinfo{booktitle}{\emph{Qualitative research design: An interactive
  approach}}. Vol.~\bibinfo{volume}{41}.
\newblock \bibinfo{publisher}{Sage publications}, \bibinfo{address}{Thousand
  Oaks, California, USA}.
\newblock


\bibitem[\protect\citeauthoryear{Navalpakkam and Churchill}{Navalpakkam and
  Churchill}{2012}]%
        {Navalpakkam.2012}
\bibfield{author}{\bibinfo{person}{Vidhya Navalpakkam} {and}
  \bibinfo{person}{Elizabeth Churchill}.} \bibinfo{year}{2012}\natexlab{}.
\newblock \showarticletitle{Mouse tracking}. In
  \bibinfo{booktitle}{\emph{Proceedings of the 2012 {ACM} annual conference on
  Human Factors in Computing Systems - {CHI} {\textquotesingle}12}}.
  \bibinfo{publisher}{{ACM} Press}, \bibinfo{address}{New York, NY, United
  States}, \bibinfo{pages}{2963--2972}.
\newblock
\urldef\tempurl%
\url{https://doi.org/10.1145/2207676.2208705}
\showDOI{\tempurl}


\bibitem[\protect\citeauthoryear{Nebeling, Speicher, and Norrie}{Nebeling
  et~al\mbox{.}}{2013}]%
        {Nebeling.2013}
\bibfield{author}{\bibinfo{person}{Michael Nebeling},
  \bibinfo{person}{Maximilian Speicher}, {and} \bibinfo{person}{Moira~C.
  Norrie}.} \bibinfo{year}{2013}\natexlab{}.
\newblock \showarticletitle{{CrowdStudy}}. In
  \bibinfo{booktitle}{\emph{Proceedings of the 5th {ACM} {SIGCHI} symposium on
  Engineering interactive computing systems - {EICS} {\textquotesingle}13}}.
  \bibinfo{publisher}{{ACM} Press}, \bibinfo{address}{London, UK},
  \bibinfo{pages}{255 -- 264}.
\newblock
\urldef\tempurl%
\url{https://doi.org/10.1145/2494603.2480303}
\showDOI{\tempurl}


\bibitem[\protect\citeauthoryear{Nielsen}{Nielsen}{1992}]%
        {Nielsen.1992}
\bibfield{author}{\bibinfo{person}{Jakob Nielsen}.}
  \bibinfo{year}{1992}\natexlab{}.
\newblock \showarticletitle{{The usability engineering life cycle}}.
\newblock \bibinfo{journal}{\emph{{Computer}}} \bibinfo{volume}{25},
  \bibinfo{number}{3} (\bibinfo{year}{1992}), \bibinfo{pages}{12--22}.
\newblock
\showISSN{1558-0814}
\urldef\tempurl%
\url{https://doi.org/10.1109/2.121503}
\showDOI{\tempurl}


\bibitem[\protect\citeauthoryear{Orlovska, Novakazi, Lars-Ola, Karlsson,
  Wickman, and Söderberg}{Orlovska et~al\mbox{.}}{2020}]%
        {Orlovska.2020}
\bibfield{author}{\bibinfo{person}{Julia Orlovska}, \bibinfo{person}{Fjollë
  Novakazi}, \bibinfo{person}{Blig{\aa}rd Lars-Ola}, \bibinfo{person}{MariAnne
  Karlsson}, \bibinfo{person}{Casper Wickman}, {and} \bibinfo{person}{Rikard
  Söderberg}.} \bibinfo{year}{2020}\natexlab{}.
\newblock \showarticletitle{Effects of the driving context on the usage of
  Automated Driver Assistance Systems ({ADAS}) -Naturalistic Driving Study for
  {ADAS} evaluation}.
\newblock \bibinfo{journal}{\emph{Transportation Research Interdisciplinary
  Perspectives}}  \bibinfo{volume}{4} (\bibinfo{date}{feb}
  \bibinfo{year}{2020}), \bibinfo{pages}{100093}.
\newblock
\urldef\tempurl%
\url{https://doi.org/10.1016/j.trip.2020.100093}
\showDOI{\tempurl}


\bibitem[\protect\citeauthoryear{Orlovska, Wickman, and S{\"o}derberg}{Orlovska
  et~al\mbox{.}}{2018a}]%
        {Orlovska.2018b}
\bibfield{author}{\bibinfo{person}{Julia Orlovska}, \bibinfo{person}{Casper
  Wickman}, {and} \bibinfo{person}{Rikard S{\"o}derberg}.}
  \bibinfo{year}{2018}\natexlab{a}.
\newblock \showarticletitle{{Big Data Analysis as a new Approach for Usability
  Attributes Evaluation of User Interfaces An Automotive Industry Context}}. In
  \bibinfo{booktitle}{\emph{{Proceedings of the DESIGN 2018 15th International
  Design Conference}}} \emph{(\bibinfo{series}{{Design Conference
  Proceedings}})}. \bibinfo{publisher}{{Faculty of Mechanical Engineering and
  Naval Architecture, University of Zagreb, Croatia} and {The Design Society,
  Glasgow, UK}}, \bibinfo{address}{Dubrovnik, Croatia},
  \bibinfo{pages}{1651--1662}.
\newblock
\urldef\tempurl%
\url{https://doi.org/10.21278/idc.2018.0243}
\showDOI{\tempurl}


\bibitem[\protect\citeauthoryear{Orlovska, Wickman, and S{\"o}derberg}{Orlovska
  et~al\mbox{.}}{2018b}]%
        {Orlovska.2018}
\bibfield{author}{\bibinfo{person}{Julia Orlovska}, \bibinfo{person}{Casper
  Wickman}, {and} \bibinfo{person}{Rikard S{\"o}derberg}.}
  \bibinfo{year}{2018}\natexlab{b}.
\newblock \showarticletitle{{Big Data Usage Can Be a Solution for User Behavior
  Evaluation: An Automotive Industry Example}}.
\newblock \bibinfo{journal}{\emph{{Procedia CIRP}}}  \bibinfo{volume}{72}
  (\bibinfo{year}{2018}), \bibinfo{pages}{117--122}.
\newblock
\showISSN{2212-8271}
\urldef\tempurl%
\url{https://doi.org/10.1016/j.procir.2018.03.102}
\showDOI{\tempurl}


\bibitem[\protect\citeauthoryear{Pfleging, Schneegass, and Schmidt}{Pfleging
  et~al\mbox{.}}{2012}]%
        {Pfleging.2012}
\bibfield{author}{\bibinfo{person}{Bastian Pfleging}, \bibinfo{person}{Stefan
  Schneegass}, {and} \bibinfo{person}{Albrecht Schmidt}.}
  \bibinfo{year}{2012}\natexlab{}.
\newblock \showarticletitle{Multimodal interaction in the car}. In
  \bibinfo{booktitle}{\emph{Proceedings of the 4th International Conference on
  Automotive User Interfaces and Interactive Vehicular Applications -
  {AutomotiveUI} {\textquotesingle}12}}. \bibinfo{publisher}{{ACM} Press},
  \bibinfo{address}{Portsmouth, New Hampshire}, \bibinfo{pages}{155--162}.
\newblock
\urldef\tempurl%
\url{https://doi.org/10.1145/2390256.2390282}
\showDOI{\tempurl}


\bibitem[\protect\citeauthoryear{Riener, Jeon, Alvarez, and Frison}{Riener
  et~al\mbox{.}}{2017}]%
        {Riener.2017}
\bibfield{author}{\bibinfo{person}{Andreas Riener}, \bibinfo{person}{Myounghoon
  Jeon}, \bibinfo{person}{Ignacio Alvarez}, {and} \bibinfo{person}{Anna~K.
  Frison}.} \bibinfo{year}{2017}\natexlab{}.
\newblock \showarticletitle{Driver in the Loop: Best Practices in Automotive
  Sensing and Feedback Mechanisms}.
\newblock In \bibinfo{booktitle}{\emph{Automotive User Interfaces}}.
  \bibinfo{publisher}{Springer International Publishing},
  \bibinfo{address}{Cham, Switzerland}, \bibinfo{pages}{295--323}.
\newblock
\urldef\tempurl%
\url{https://doi.org/10.1007/978-3-319-49448-7_11}
\showDOI{\tempurl}


\bibitem[\protect\citeauthoryear{Risteska, Chakraborty, and Donmez}{Risteska
  et~al\mbox{.}}{2018}]%
        {Risteska.2018}
\bibfield{author}{\bibinfo{person}{Martina Risteska}, \bibinfo{person}{Joyita
  Chakraborty}, {and} \bibinfo{person}{Birsen Donmez}.}
  \bibinfo{year}{2018}\natexlab{}.
\newblock \showarticletitle{Predicting Environmental Demand and Secondary Task
  Engagement using Vehicle Kinematics from Naturalistic Driving Data}. In
  \bibinfo{booktitle}{\emph{Proceedings of the 10th International Conference on
  Automotive User Interfaces and Interactive Vehicular Applications -
  {AutomotiveUI} {\textquotesingle}18}}. \bibinfo{publisher}{{ACM} Press},
  \bibinfo{address}{New York, NY, United States}, \bibinfo{pages}{66--73}.
\newblock
\urldef\tempurl%
\url{https://doi.org/10.1145/3239060.3239091}
\showDOI{\tempurl}


\bibitem[\protect\citeauthoryear{Roto, Law, Vermeeren, and Hoonhout}{Roto
  et~al\mbox{.}}{2011}]%
        {Roto.2011}
\bibfield{author}{\bibinfo{person}{Virpi Roto}, \bibinfo{person}{Effie
  Lai-Chong Law}, \bibinfo{person}{Arnold P. O.~S. Vermeeren}, {and}
  \bibinfo{person}{Jettie Hoonhout}.} \bibinfo{year}{2011}\natexlab{}.
\newblock \showarticletitle{User Experience White Paper – Bringing clarity to
  the concept of user experience}. In \bibinfo{booktitle}{\emph{Dagstuhl
  Seminar on Demarcating User Experience}}. \bibinfo{publisher}{Dagstuhl
  reports}, \bibinfo{address}{Dagstuhl, Germany}, \bibinfo{pages}{1--26}.
\newblock


\bibitem[\protect\citeauthoryear{Roto, Rantavuo, and V{\"a}{\"a}n{\"a}nen}{Roto
  et~al\mbox{.}}{2009}]%
        {Roto.2009}
\bibfield{author}{\bibinfo{person}{Virpi Roto}, \bibinfo{person}{Heli
  Rantavuo}, {and} \bibinfo{person}{Kaisa V{\"a}{\"a}n{\"a}nen}.}
  \bibinfo{year}{2009}\natexlab{}.
\newblock \showarticletitle{{Evaluating user experience of early product
  concepts}}.
\newblock \bibinfo{journal}{\emph{{Proceeding of the International Conference
  on Designing Pleasurable Products and Interfaces DPPI09}}}
  \bibinfo{volume}{4} (\bibinfo{year}{2009}), \bibinfo{pages}{1--10}.
\newblock


\bibitem[\protect\citeauthoryear{Salda{\~n}a}{Salda{\~n}a}{2015}]%
        {Saldana.2015}
\bibfield{author}{\bibinfo{person}{Johnny Salda{\~n}a}.}
  \bibinfo{year}{2015}\natexlab{}.
\newblock \bibinfo{booktitle}{\emph{{The coding manual for qualitative
  researchers}}}.
\newblock \bibinfo{publisher}{Sage}, \bibinfo{address}{New York, NY, United
  States}.
\newblock


\bibitem[\protect\citeauthoryear{Schneegass, Pfleging, Broy, Schmidt, and
  Heinrich}{Schneegass et~al\mbox{.}}{2013}]%
        {Schneegass.2013}
\bibfield{author}{\bibinfo{person}{Stefan Schneegass}, \bibinfo{person}{Bastian
  Pfleging}, \bibinfo{person}{Nora Broy}, \bibinfo{person}{Albrecht Schmidt},
  {and} \bibinfo{person}{Frederik Heinrich}.} \bibinfo{year}{2013}\natexlab{}.
\newblock \showarticletitle{A data set of real world driving to assess driver
  workload}. In \bibinfo{booktitle}{\emph{Proceedings of the 5th International
  Conference on Automotive User Interfaces and Interactive Vehicular
  Applications - {AutomotiveUI} {\textquotesingle}13}}.
  \bibinfo{publisher}{{ACM} Press}, \bibinfo{address}{Eindhove, Netherlands},
  \bibinfo{pages}{150--157}.
\newblock
\urldef\tempurl%
\url{https://doi.org/10.1145/2516540.2516561}
\showDOI{\tempurl}


\bibitem[\protect\citeauthoryear{Schroeter, Rakotonirainy, and Foth}{Schroeter
  et~al\mbox{.}}{2012}]%
        {Schroeter.2012}
\bibfield{author}{\bibinfo{person}{Ronald Schroeter}, \bibinfo{person}{Andry
  Rakotonirainy}, {and} \bibinfo{person}{Marcus Foth}.}
  \bibinfo{year}{2012}\natexlab{}.
\newblock \showarticletitle{The social car}. In
  \bibinfo{booktitle}{\emph{Proceedings of the 4th International Conference on
  Automotive User Interfaces and Interactive Vehicular Applications -
  {AutomotiveUI} {\textquotesingle}12}}. \bibinfo{publisher}{{ACM} Press},
  \bibinfo{address}{Portsmouth, New Hampshire}, \bibinfo{pages}{107--110}.
\newblock
\urldef\tempurl%
\url{https://doi.org/10.1145/2390256.2390273}
\showDOI{\tempurl}


\bibitem[\protect\citeauthoryear{Speicher, Both, and Gaedke}{Speicher
  et~al\mbox{.}}{2015}]%
        {Speicher.2015}
\bibfield{author}{\bibinfo{person}{Maximilian Speicher},
  \bibinfo{person}{Andreas Both}, {and} \bibinfo{person}{Martin Gaedke}.}
  \bibinfo{year}{2015}\natexlab{}.
\newblock \showarticletitle{Inuit: The Interface Usability Instrument}.
\newblock In \bibinfo{booktitle}{\emph{Design, User Experience, and Usability:
  Design Discourse}}. \bibinfo{publisher}{Springer International Publishing},
  \bibinfo{address}{Heidelberg, Germany}, \bibinfo{pages}{256--268}.
\newblock
\urldef\tempurl%
\url{https://doi.org/10.1007/978-3-319-20886-2_25}
\showDOI{\tempurl}


\bibitem[\protect\citeauthoryear{Väänänen, Roto, and Hassenzahl}{Väänänen
  et~al\mbox{.}}{2008}]%
        {Vaananen.2008}
\bibfield{author}{\bibinfo{person}{Kaisa Väänänen}, \bibinfo{person}{Virpi
  Roto}, {and} \bibinfo{person}{Marc Hassenzahl}.}
  \bibinfo{year}{2008}\natexlab{}.
\newblock \showarticletitle{Towards Practical User Experience Evaluation
  Methods}.
\newblock \bibinfo{journal}{\emph{Meaningful measures: Valid useful user
  experience measurement (VUUM)}}  \bibinfo{volume}{{}} (\bibinfo{date}{01}
  \bibinfo{year}{2008}), \bibinfo{pages}{1--4}.
\newblock


\bibitem[\protect\citeauthoryear{Wang, Zhang, Tang, Zheng, and Zhao}{Wang
  et~al\mbox{.}}{2016}]%
        {Wang.2016}
\bibfield{author}{\bibinfo{person}{Gang Wang}, \bibinfo{person}{Xinyi Zhang},
  \bibinfo{person}{Shiliang Tang}, \bibinfo{person}{Haitao Zheng}, {and}
  \bibinfo{person}{Ben~Y. Zhao}.} \bibinfo{year}{2016}\natexlab{}.
\newblock \showarticletitle{{Unsupervised Clickstream Clustering for User
  Behavior Analysis}}. In \bibinfo{booktitle}{\emph{{Proceedings of the 2016
  CHI Conference on Human Factors in Computing Systems}}}
  \emph{(\bibinfo{series}{{CHI '16}})}. \bibinfo{publisher}{{Association for
  Computing Machinery}}, \bibinfo{address}{New York, NY, USA},
  \bibinfo{pages}{225--236}.
\newblock
\showISBNx{9781450333627}
\urldef\tempurl%
\url{https://doi.org/10.1145/2858036.2858107}
\showDOI{\tempurl}


\bibitem[\protect\citeauthoryear{Zhang, Patel, Buthpitiya, Lyons, Harrison, and
  Abowd}{Zhang et~al\mbox{.}}{2016}]%
        {Zhang.2016}
\bibfield{author}{\bibinfo{person}{Cheng Zhang}, \bibinfo{person}{Mitesh
  Patel}, \bibinfo{person}{Senaka Buthpitiya}, \bibinfo{person}{Kent Lyons},
  \bibinfo{person}{Beverly Harrison}, {and} \bibinfo{person}{Gregory~D.
  Abowd}.} \bibinfo{year}{2016}\natexlab{}.
\newblock \showarticletitle{Driver Classification Based on Driving Behaviors}.
  In \bibinfo{booktitle}{\emph{Proceedings of the 21st International Conference
  on Intelligent User Interfaces - {IUI} {\textquotesingle}16}}.
  \bibinfo{publisher}{{ACM} Press}, \bibinfo{address}{New York, NY, United
  States}, \bibinfo{pages}{80--84}.
\newblock
\urldef\tempurl%
\url{https://doi.org/10.1145/2856767.2856806}
\showDOI{\tempurl}


\end{thebibliography}

\end{document}